# Min-Bias and the Underlying Event at the LHC

Rick Field
Department of Physics, University of Florida, Gainesville, FL 32611, USA

The CDF PYTHIA 6.2 Tune DW predictions for the behavior of the LHC underlying event (UE) data at 900 GeV and 7 TeV are examined in detail. The behavior of the UE at the LHC is roughly what we expected. The LHC PYTHIA 6.4 Tune Z1 does an excellent job describing the UE data at 900 GeV and 7 TeV. Both Tune DW and Tune Z1 describe fairly well the Drell-Yan UE data at 1.96 TeV and 7 TeV. No model describes all the features of "min-bias" (MB) collisions at 900 GeV and 7 TeV.

## 1. INTRODUCTION

The total proton-proton cross section is the sum of the elastic and inelastic components, $\sigma_{tot} = \sigma_{EL} + \sigma_{INEL}$. Elastic scattering is a 2-to-2 color singlet exchange process in which the two outgoing particles are the same as the two incoming particles. Single and double-diffraction also corresponds to color singlet exchange between the initial hadrons. For single-diffraction (SD) one of the incoming protons and double-diffraction (DD) both of the incoming protons are excited into a high mass color singlet state (*i.e.* N* states) which then decays. When color is exchanged the outgoing remnants are no longer color singlets and one has a separation of color resulting in a multitude of quark-antiquark pairs being pulled out of the vacuum. The "non-diffractive" (ND) component involves color exchange and the separation of color. However, the "non-diffractive" collisions have both a "soft" and "hard" component. Most of the time the color exchange between partons in the beam hadrons occurs through a soft interaction (*i.e.* no high transverse momentum) and the two beam hadrons "ooze" through each other producing lots of soft particles with a uniform distribution in rapidity and many particles flying down the beam pipe. Occasionally, there is a hard scattering among the constituent partons producing outgoing particles and "jets" with high transverse momentum. In order to simulate an inelastic hadron-hadron collision one must not only be able to model the ND component, but also SD and DD.

Min-bias (MB) is a generic term which refers to events that are selected with a "loose" trigger that accepts a large fraction of the overall inelastic cross section. All triggers produce some bias and the term "min-bias" is meaningless until one specifies the precise trigger used to collect the data. The underlying event (UE) consists of particles that accompany a hard scattering such as the beam-beam remnants (BBR) and particles originating from multiple-parton interactions (MPI). Initial and final-state radiation can also contribute particles to the UE. MB and UE are not the same object! The majority of MB collisions are "soft", while the UE is studied in events in which a hard-scattering has occurred. In the traditional approach, one uses the topological structure of the hard hadron-hadron collision to experimentally study the UE [1, 2]. On an event-by-event bases one selects a "leading object" which is used to define three regions of $\eta$-$\phi$ space. The pseudo-rapidity is defined by $\eta = -\log(\tan(\theta_{cm}/2))$, where $\theta_{cm}$ is the center-of-mass polar scattering angle and $\phi$ is the azimuthal angle of outgoing particles. The angle $\Delta\phi = \phi - \phi_1$ is the relative azimuthal angle between charged particles, $\phi$, and the direction "leading object", $\phi_1$. The "toward" region is defined by $|\Delta\phi| < 60°$ and $|\eta| < \eta_{cut}$, while the "away" region is $|\Delta\phi| > 120°$ and $|\eta| < \eta_{cut}$. The "transverse" region $60° < |\Delta\phi| < 120°$ and $|\eta| < \eta_{cut}$ is roughly perpendicular to the plane of the hard 2-to-2 parton-parton scattering and is therefore very sensitive to the modeling of the UE. Each of the three regions has an area in $\eta$-$\phi$ space of $\Delta\eta\Delta\phi = 2\eta_{cut} \times 2\pi/3$. For example, the "transverse" charged particle density is the number of charged particles with $p_T > $ PTmin in the "transverse" region divided by the area in $\eta$-$\phi$ space. Similarly, the "transverse" charged PTsum density is the *scalar* PTsum of charged particles with $p_T > $ PTmin in the "transverse" region divided by the area in $\eta$-$\phi$ space. The "leading object" can by the leading (highest $p_T$) calorimeter jet, jet#1, or the leading charged particle jet, chgjet#1, or the leading charged particle, PTmax. For Drell-Yan production the "leading object" is the lepton-pair and if the mass of the lepton-pair is near the Z-boson then the "leading object" is the Z-boson. For Drell-Yan production the lepton-pair are not included in the charged particle density or the PTsum density.

Figure 1 shows the charged particle density in the "transverse" region for charged particles ($p_T > 0.5$ GeV/c, $|\eta| < 1$) at 7 TeV as defined by the leading charged particle, PTmax, the leading charged particle jet, chgjet#1, and the muon-pair in Z-boson production as predicted from PYTHIA 6.2 Tune DW. The density of charged particles in the "transverse" region goes to zero as PTmax or PT(chgjet#1) go to zero due to kinematics. If PTmax is equal to zero then there are no charged particles anywhere in the $\eta$ region considered. Similarly for PT(chgjet#1). However, if PT(muon-pair) goes to zero there is still the hard scale of the mass of the Z-boson and, hence, the charge particle density is not zero at PT(muon-pair) = 0.

QCD Monte-Carlo generators such as PYTHIA [3] have parameters which may be adjusted to control the behavior of their event modeling. A specified set of these parameters that has been adjusted to better fit some aspects of the data is referred to as a tune [4, 5]. PYTHIA 6.2 Tune DW [6] is a CDF Run 2 tune that does a very nice job in describing the CDF UE data. PYTHIA 6.4 Tune Z1 [7] is a CMS LHC tune that fits very well the LHC UE data at 900 GeV and 7 TeV. In Section 2, I will compare Tune DW and Tune Z1 with the Tevatron and LHC UE data. In Section 3, we will examine how well the tunes do in describing the LHC MB data. Section 4 contains a short summary plus some conclusions.

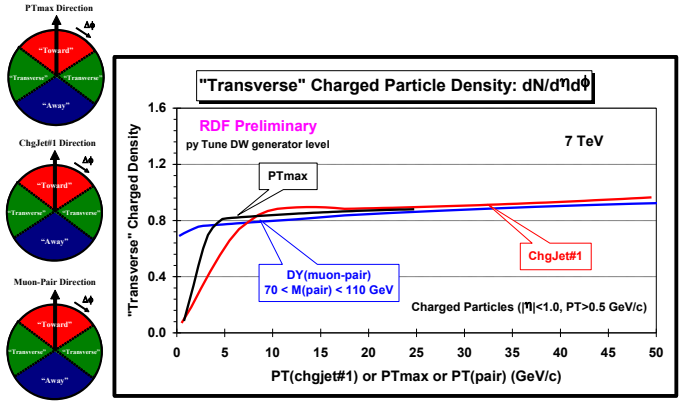

Figure 1: Shows the charged particle density in the "transverse" region for charged particles ($p_T > 0.5$ GeV/c, $|\eta| < 1$) at 7 TeV as defined by the leading charged particle, PTmax, the leading charged particle jet, chgjet#1, and the muon-pair in Z-boson production as predicted from PYTHIA 6.2 Tune DW at the particle level. For Z-boson production the muon-pair are excluded from the charged particle density. Charged particle jets are constructed using the Anti-KT algorithm with d = 0.5.

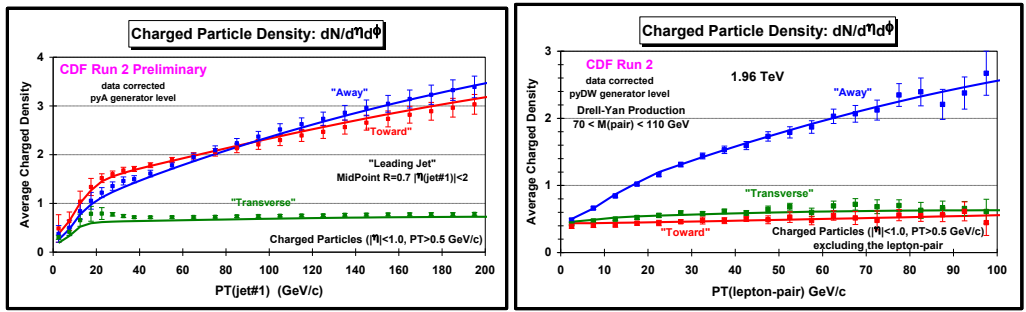

Figure 2: CDF data at 1.96 TeV [6] on the density of charged particles, $dN/d\eta d\phi$, with $p_T > 0.5$ GeV/c and $|\eta| < 1$ for "leading jet" (*left*) and "Z-boson" (*right*) events as a function of the leading jet $p_T$ and $p_T$(lepton-pair), respectively, for the "toward", "away", and "transverse" regions. The data are corrected to the particle level and are compared with PYTHIA 6.2 Tune A and Tune AW, respectively, at the particle level [4].

## 2. THE UE AT THE TEVATRON AND THE LHC

Figure 2 shows CDF data at 1.96 TeV on the density of charged particles and the *scalar* PTsum density, respectively, for the "toward", "away", and "transverse" regions for "leading jet" and "Z-boson" events. For "leading jet" events the densities are plotted as a function of the leading jet $p_T$ and for "Z-boson" events there are plotted versus $p_T(Z)$. The data are corrected to the particle level and are compared with PYTHIA Tune A ("leading jet") and Tune AW ("Z-boson") at the particle level. For "leading jet" events at high $p_T$(jet#1) the densities in the "toward" and "away" regions are much larger than in the "transverse" region because of the "toward-side" and "away-side" jets. At small $p_T$(jet#1) the "toward", "away", and "transverse" densities become equal and go to zero as $p_T$(jet#1) goes to zero. For "Z-boson" events the "toward" and "transverse" densities are both small and almost equal. The "away" density is large due to the "away-side" jet. The "toward", "away", and "transverse" densities become equal as $p_T(Z)$ goes to zero, but unlike the "leading jet" case the densities do not vanish at $p_T(Z) = 0$.

Figure 3 compares the CDF data at 1.96 TeV with the recent CMS data at 7 TeV on the density of charged particles, "toward", "away", and "transverse" regions for Z-boson production. For CDF the charged particles have $p_T > 0.5$ GeV/c and $|\eta| < 1$; the leptons have $p_T > 20$ GeV and $|\eta| < 1.0$ and are not included in the charged particle density and the lepton-pair is required to have $70 < M$(lepton-pair) $< 110$ GeV. For CMS the charged particles have $p_T > 0.5$ GeV/c and $|\eta| < 2$; the leptons have $p_T > 20$ GeV and $|\eta| < 2.4$ and are not included in the charged particle density and the lepton-pair is required to have 60 < M(lepton-pair) < 120 GeV. The data are corrected to the particle level and are compared with Tune DW and Tune Z1 at the particle level.

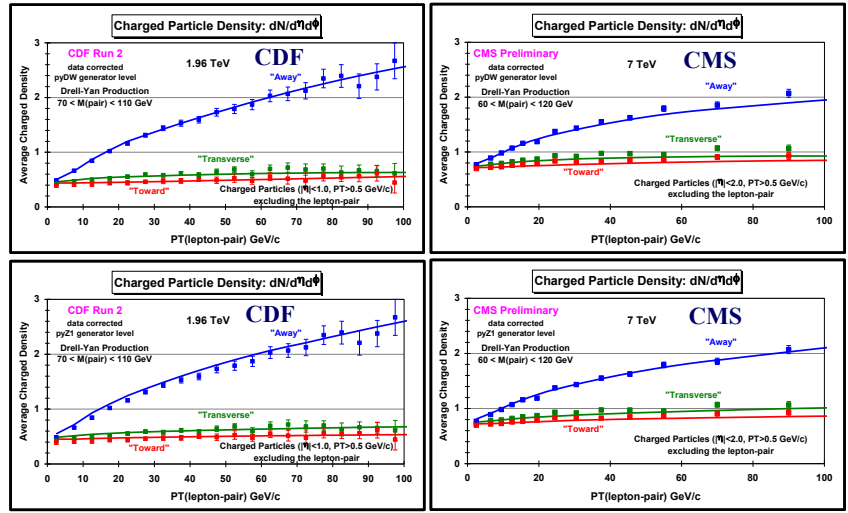

Figure 3: (*left column*) CDF data at 1.96 TeV [6] on the density of charged particles, dN/dηdφ, with $p_T$ > 0.5 GeV/c and |η| < 1 for "Z-boson" events as a function of the $p_T$(lepton-pair) for the "toward", "away", and "transverse" regions. The leptons are required to have $p_T$ > 20 GeV and |η| < 1.0 and are not included in the charged particle density. The lepton-pair is required to have 70 < M(lepton-pair) < 110 GeV. (*right column*) Recent CMS data [8] at 7 TeV on the density of charged particles, dN/dηdφ, with $p_T$ > 0.5 GeV/c and |η| < 2 for "Z-boson" events as a function of the $p_T$(lepton-pair) for the "toward", "away", and "transverse" regions. The leptons are required to have $p_T$ > 20 GeV and |η| < 2.4 and are not included in the charged particle density. The lepton-pair is required to have 60 < M(lepton-pair) < 120 GeV. The data are corrected to the particle level and are compared with PYTHIA 6.2 Tune DW (*top row*) and PYTHIA 6.4 Tune Z1 (*bottom row*) at the particle level.

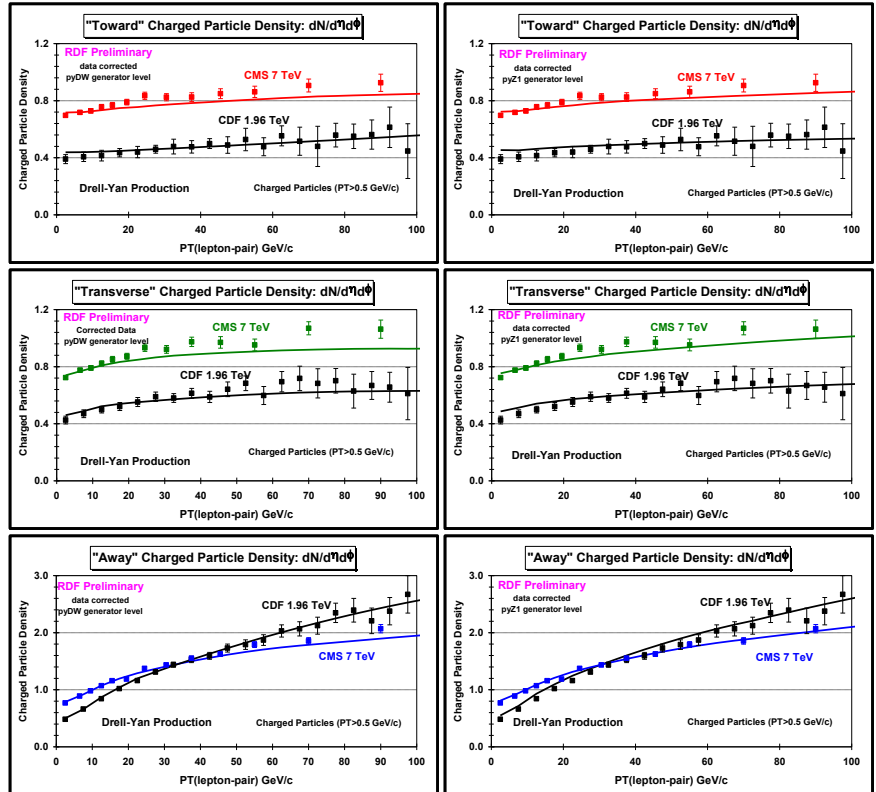

Figure 4: Compares the CDF data and the CMS data from Fig. 3 on the density of charged particles, dN/dηdφ, for "Z-boson" events as a function of the $p_T$(lepton-pair) for the "toward", "away", and "transverse" regions. The data are corrected to the particle level and are compared with PYTHIA 6.2 Tune DW (*left column*) and PYTHIA 6.4 Tune Z1 (*right column*) at the particle level.

Figure 4 shows detailed comparisons of the CDF data at 1.96 TeV and the CMS data at 7 TeV for "Z-boson" events. The large increase in the activity in the "toward" and "transverse" regions in going from 1.96 TeV to 7 TeV was expected due to the increased MPI at 7 TeV [6]. Both Tune DW and Tune Z1 describe this increase fairly well. The "away" region is affected by the different lepton cuts used by CDF and CMS, however, both tunes also describe this region fairly well.

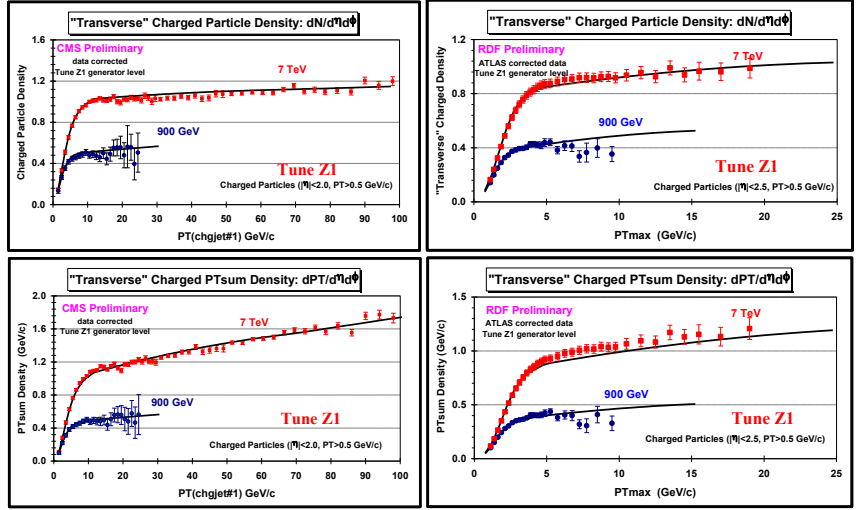

Figure 5: (*left column*) CMS data [9] at 900 GeV and 7 TeV on the transverse charged particle density (*top left*) and the transverse charged PTsum density (*bottom left*) as defined by the leading charged particle jet (chgjet#1) for charged particles with $p_T > 0.5$ GeV/c and $|\eta| < 2$. (*right column*) ATLAS data [10] at 900 GeV and 7 TeV on the transverse charged particle density (*top right*) and the transverse charged PTsum density (*bottom right*) as defined by the leading charged particle, PTmax, as a function of PTmax for charged particles with $p_T > 0.5$ GeV/c and $|\eta| < 2.5$. The data are corrected to the particle level and compared with PYTHIA 6.4 Tune Z1 at the generator level.

Figure 5 shows the recent CMS data at 900 GeV and 7 TeV on the transverse charged particle density and the transverse charged PTsum density as defined by the leading charged particle jet (chgjet#1) for charged particles with $p_T > 0.5$ GeV/c and $|\eta| < 2$. Figure 5 also shows the recent ATLAS data at 900 GeV and 7 TeV on the transverse charged particle density and the transverse charged PTsum density as defined by the leading charged particle, PTmax, as a function of PTmax for charged particles with $p_T > 0.5$ GeV/c and $|\eta| < 2.5$. The CMS and ATLAS data are corrected to the particle level and compared with Tune Z1 at the generator level. Tune Z1 does a very nice job in describing both the CMS and ATLAS UE data.

Figure 6 shows the CDF data at 1.96 TeV on the charged particle density in the "transverse" region as defined by the leading calorimeter jet, jet#1, together with the CMS data at 900 GeV and 7 TeV on the charged particle density in the "transverse" region as defined by the leading charged particle jet, chgjet#1, compared with PYTHIA Tune DW and Tune Z1. Both tunes do a fairy good job in describing all three energies. Tune DW, however, is not a perfect fit to the LHC UE data. It does not fit the Tevatron perfectly either! Tune Z1 is in very good agreement with the UE data at 900 GeV and 7 TeV but a little high at 1.96 TeV. We expect a lot from the QCD Monte-Carlo models. We want them to fit everything perfectly which is, of course, not always possible. Nevertheless I believe that we will find a PYTHIA tune that simultaneously describes 900 GeV, 1.96 TeV, and 7 TeV. Remember the energy dependence of the UE in PYTHIA depends not only on $\varepsilon$ = PARP (90), but also on the choice of PDF [11].

Figure 7 compares the CMS data using chgjet#1 with the ATLAS data which uses the PTmax approach. Tune Z1 describes the differences between the CMS chgjet#1 and the ATLAS PTmax approach very well. It is interesting that the activity in the "transverse" region in the "plateau" is larger for the chgjet#1 analysis than it is for the PTmax analysis. Could it be that when one requires a charged particle jet with a certain value of PT(chgjet#1) that you bias the UE to be more active, because a more active UE can contribute some $p_T$ to the leading charged particle jet? In an attempt to understand this, in Fig. 7 I looked at the dependence of the transverse charged particle density on the charged particle jet size (*i.e.* radius) as predicted by PYTHIA Tune Z1. I constructed charged particle jets using the Anti-KT algorithm with d = 0.2, 0.5, and 1.0. The charged particles have $p_T > 0.5$ GeV/c and $|\eta| < 2.5$ and the leading charged particle jet is restricted to be in the region $|\eta(\text{chgjet\#1})| < 1.5$. For very narrow jets the UE "plateau" is nearly the same as in the PTmax approach. As the jets become large in radius, the UE "plateau" becomes more active! The object that is being used to define the "transverse" region can indeed bias the UE to be more active. Amazing!

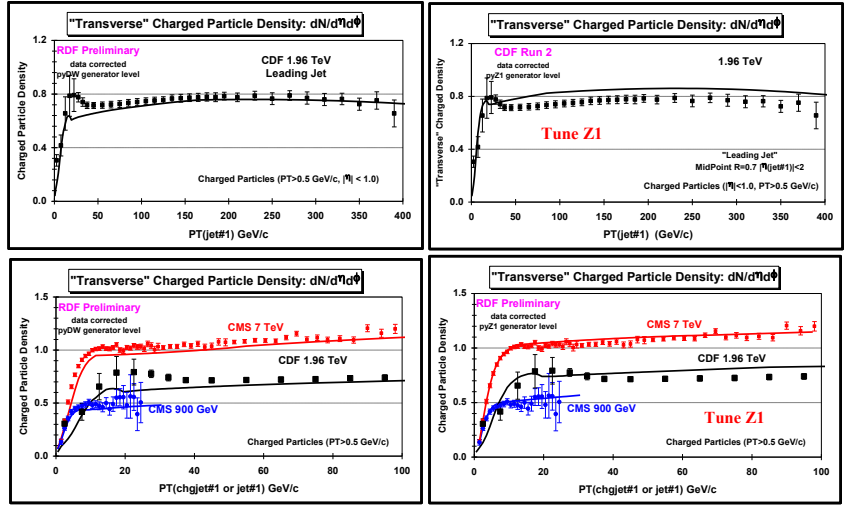

Figure 6: (*top row*) CDF data at 1.96 TeV [6] on the charged particle density ($p_T > 0.5$ GeV/c, $|\eta| < 1$) in the "transverse" region as defined by the leading calorimeter jet, jet#1, as a function of PT(jet#1) (*bottom row*) Compares the CDF data at 1.96 TeV with the CMS data [9] at 900 GeV and 7 TeV on the "transverse" charged particle density ($p_T > 0.5$ GeV/c, $|\eta| < 2$) in the "transverse" region as defined by the leading charged particle jet, chgjet#1, as a function of PT(chgjet#1). The data are corrected to the particle level and compared with PYTHIA 6.2 Tune DW (*left column*) and PYTHIA 6.4 Tune Z1 (*right column*) at the generator level.

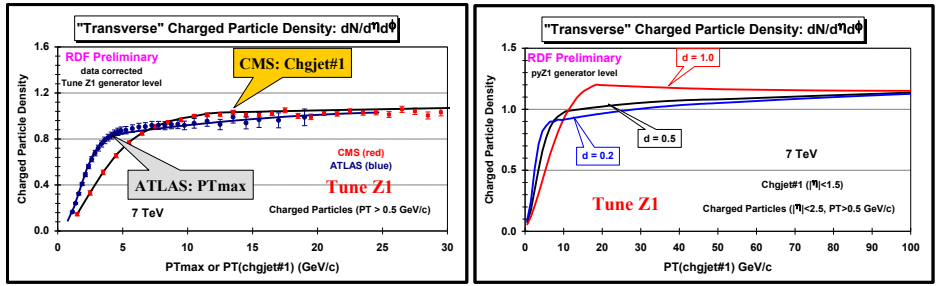

Figure 7: (*left*) CMS data from Fig 5 on the charged particle density in the "transverse" region as defined by the leading charged particle jet, chgjet#1, versus PT(chgjet#1) compared with the ATLAS data from Fig. 5 on the charged particle density in the "transverse" region as defined by the leading charged particle, PTmax, versus PTmax. The data are corrected to the particle level and compared with PYTHIA 6.4 Tune Z1 at the generator level. (*right*) Dependence of the transverse charged particle density on the charged particle jet size (*i.e.* radius) as predicted by Tune Z1. Charged particle jets are constructed using the Anti-KT algorithm with d = 0.2, 0.5, and 1.0. The charged particles have $p_T > 0.5$ GeV/c and $|\eta| < 2.5$ the leading charged particle jet is restricted to be in the region $|\eta(\text{chgjet\#1})| < 1.5$.

## 3. PREDICTING MB AT THE LHC

The perturbative 2-to-2 parton-parton differential cross section diverges like $1/\hat{p}_T^4$, where $\hat{p}_T$ is the transverse momentum of the outgoing parton in the parton-parton center-of-mass frame. PYTHIA regulates this cross section by including a smooth cut-off $p_{T0}$ as follows: $1/\hat{p}_T^4 \to 1/(\hat{p}_T^2 + p_{T0}^2)^2$. This approaches the perturbative result for large scales and is finite as $\hat{p}_T \to 0$. The primary hard scattering processes and the MPI are regulated in the same way with the one parameter $p_{T0}$ = PARP(82). This parameter governs the amount of MPI in the event and can be determined by fitting UE data. Smaller values of $p_{T0}$ results in more MPI due to a larger MPI cross-section. Since PYTHIA regulates both the primary hard scattering and the MPI with the same cut-off, $p_{T0}$, with PYTHIA one can model the overall "non-diffractive" (ND) cross section by simply letting the transverse momentum of the primary hard scattering go to zero (with no additional parameters). The non-diffractive cross section then consists of BBR plus "soft" MPI with one of the MPI occasionally being hard. In this simple approach the UE in a hard-scattering process is related to MB collisions, but they are not the same. Of course, to model MB collisions one must also add a model of single (SD) and double diffraction (DD). This makes the modeling of MB much more complicated than the modeling of the UE and one cannot trust the PYTHIA 6.2 modeling of SD and DD. Both PYTHIA Tune DW and Tune Z1 were determined by fitting UE data. They were not tuned to fit MB data. It is interesting to see how well they predict MB data.

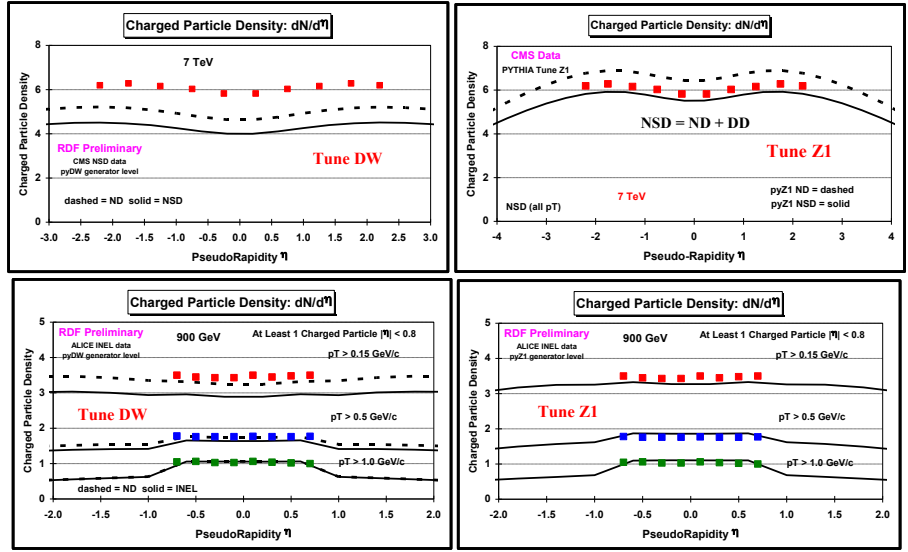

Figure 8: (*top row*) The non-single diffraction (NSD) data from CMS [12] at 7 TeV on the charged particle density, dN/dη (all $p_T$) compared with PYTHIA 6.2 Tune DW (*top left*) and PYTHIA 6.4 Tune Z1 (*top right*). The solid curve is NSD and the dashed curve is inelastic non-diffraction (ND) component. (*bottom row*) The inelastic (INEL) data from ALICE [13] at 900 GeV on the charged particle density, dN/dη, with $p_T$ > PTcut and at lease one charged particle with $p_T$ > PTcut and |η| < 0.8 for PTcut = 0.15 GeV/c, 0.5 GeV/c, and 1.0 GeV/c compared with PYTHIA 6.2 Tune DW (*bottom left*) and PYTHIA 6.4 Tune Z1 (*bottom right*). The solid curve is the INEL and the dashed curve is inelastic non-diffraction (ND) component.

Figure 8 shows the non-single diffraction (NSD) data from CMS 7 TeV on the charged particle density, dN/dη (all $p_T$) compared with PYTHIA Tune DW and Tune Z1. The solid curve is NSD and the dashed curve is inelastic non-diffraction (ND) component. The NSD cross section is the sum of ND + DD. Figure 8 also shows the INEL data from ALICE at 900 GeV on the charged particle density, dN/dη, with $p_T$ > PTcut and at lease one charged particle with $p_T$ > PTcut and |η| < 0.8 for PTcut = 0.15 GeV/c, 0.5 GeV/c, and 1.0 GeV/c compared with PYTHIA Tune DW and Tune Z1. Tune DW was tuned to fit the Tevatron data with $p_T$ > 0.5 GeV/c. Two things change when we extrapolate from the Tevatron to the LHC. Of course the center-of-mass energy changes, but also we are looking at softer particles (*i.e.* $p_T$ < 500 MeV/c). Figure 8 shows that Tune DW does okay for $p_T$ > 500 MeV/c, but does not produce enough soft particles below 500 MeV/c. One can also see that, at least in PYTHIA 6.2, the modeling of SD and DD is more important at the lower $p_T$ values. Tune Z1 does a better job at fitting the LHC MB data than does Tune DW since it produces more soft particles below 500 MeV/c than does Tune DW. However, Tune Z1 does not fit the data perfectly.

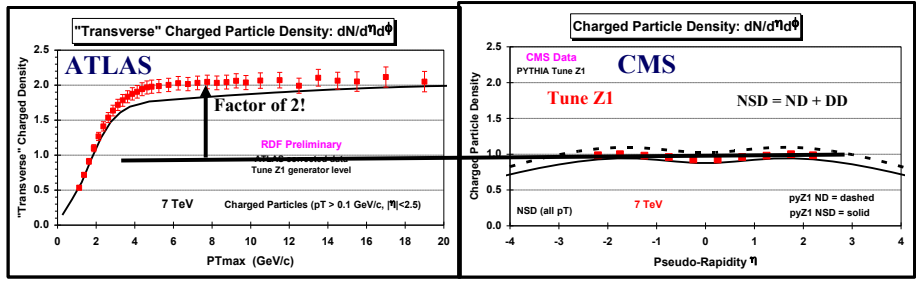

Figure 9: (*right*) The non-single diffraction (NSD) data from CMS 7 TeV on the charged particle density, dN/dηdφ (all $p_T$) compared with PYTHIA 6.4 Tune Z1. The data and theory on dN/dη in Fig. 8 has been divided by 2π to construct the number of particles per unit η-φ. (*left*) ATLAS data [10] at 7 TeV on the charged particle density in the "transverse" region as defined by the leading charged particle, PTmax, as a function of PTmax for charged particles with $p_T$ > 0.1 GeV/c and |η| < 2.5 compared with PYTHIA 6.4 Tune Z1. The activity in the UE of a hard scattering process (*left*) is a factor of two greater than it is in an average MB collision (*right*).

Figure 9 compares the activity in the UE of a hard scattering process with an average MB collision. The activity in the UE of a hard scattering process at 7 TeV is roughly a factor of two greater than it is for an average MB collision at 7 TeV and Tune Z1 describes this difference fairly well. In PYTHIA this difference comes from the fact that there are more MPI in a hard scattering process than in a typical MB collision. By demanding a hard scattering you force the collision to be more central (*i.e.* smaller impact parameter), which increases the chance of MPI.

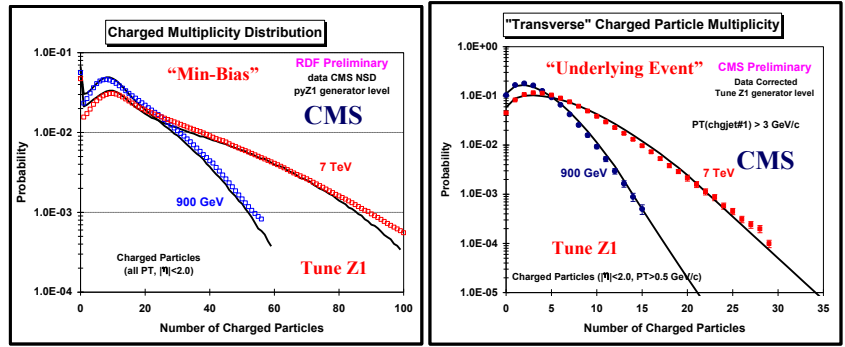

Figure 10: (*left*) The non-single diffraction (NSD) data from CMS [14] at 900 GeV and 7 TeV on the charged particle multiplicity distribution ($|\eta| < 2$, all $p_T$). (*right*) Data from CMS [9] at 900 GeV and 7 TeV on the charged particle multiplicity distribution ($|\eta| < 2$, $p_T > 0.5$ GeV/c) in the "transverse" region as defined by the leading charged particle jet, chgjet#1, for PT(chgjet#1) > 3.0 GeV/c. The data have been corrected to the particle level and compared with PYTHIA 6.4 Tune Z1 at the generator level.

Figure 10 shows the data from CMS at 900 GeV and 7 TeV on the charged particle multiplicity distribution ($|\eta| < 2$, all $p_T$) compared with PYTHIA Tune Z1 and the data from CMS at 900 GeV and 7 TeV on the charged particle multiplicity distribution in the "transverse" region as defined by the leading charged particle jet, chgjet#1, for PT(chgjet#1) > 3.0 GeV/c compared with PYTHIA Tune Z1. You are asking a lot of the QCD Monte-Carlo model when you expect it to simultaneously describe both MB and the UE in a hard scattering process. I think it is amazing that Tune Z1 does as well as it does in describing both!

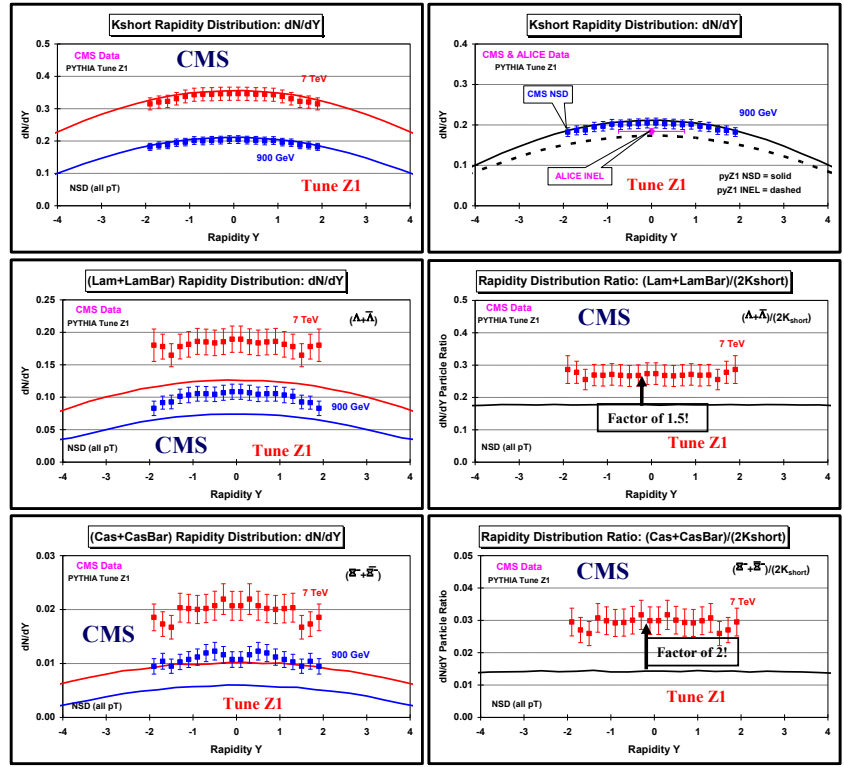

Figure 11: Recent CMS data [15] at 900 GeV and 7 TeV on the NSD rapidity distribution, dN/dY, for $K_{short}$ (*top left*), $\Lambda + \overline{\Lambda}$ (*middle left*), and $\Xi^- + \overline{\Xi}^-$ (*bottom left*). CMS data at 900 GeV and 7 TeV on the NSD rapidity distribution, dN/dY, for $K_{short}$ compared with the ALICE [16] INEL value (*top right*). The particle ratio $(\Lambda + \overline{\Lambda})/(2K_{short})$ (*middle right*) and $(\Xi^- + \overline{\Xi}^-)/(2K_{short})$ (*bottom right*) from CMS at 7 TeV. The data are compared with PYTHIA 6.4 Tune Z1.

Figure 11 shows the recent CMS data at 900 GeV and 7 TeV on the NSD rapidity distribution, dN/dY, for $K_{short}$, $\Lambda + \overline{\Lambda}$, and $\Xi^- + \overline{\Xi}^-$ compared with PYTHIA Tune Z1. There is no overall shortage of kaons in Tune Z1. For $K_{short}$ Tune Z1 is right on the CMS data at both 900 GeV and 7 TeV! However, it is a little low on the overall number of charged particles (see Fig. 8). Tune Z1 does a fairly good job in describing the difference between the ALICE INEL $K_{short}$ yield and the CMS NSD $K_{short}$ yield at 900 GeV. To predict the INEL value one must include both SD and NSD and this causes the ALICE point to lie slightly below the CMS data.

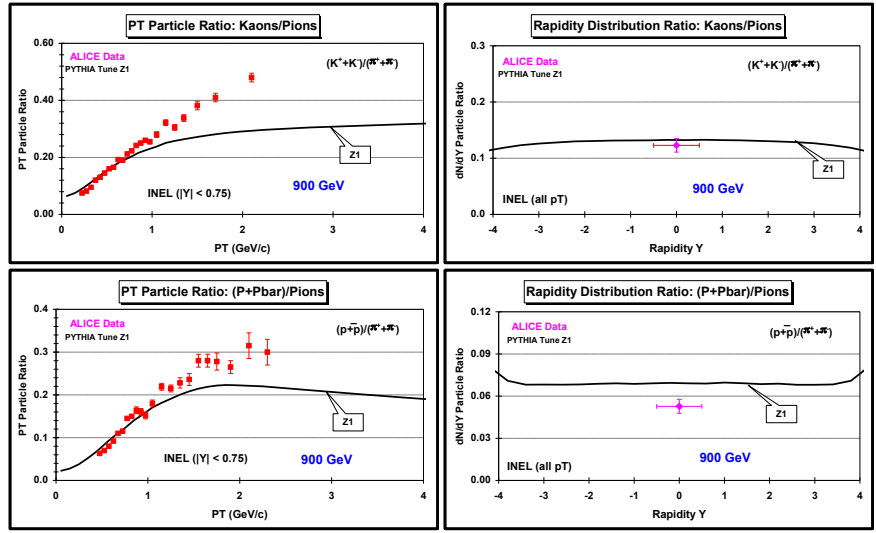

Figure 12: ALICE data [16] at 900 GeV on the INEL rapidity distribution, dN/dY, ratio for $(K^+ + K^-)/(\pi^+ + \pi^-)$ (*top right*) and $(p + \overline{p})/(\pi^+ + \pi^-)$(*bottom right*) at Y = 0 (*top left*). Also shows these ratios versus $p_T$ (*left column*). The data are compared with PYTHIA 6.4 Tune Z1.

For strange baryons it is a much different story. Tune Z1 does not produce as many strange baryons as are seen in the data at both 900 GeV and 7 TeV. Tune Z1 is low by a factor of about 1.5 for $\Lambda + \overline{\Lambda}$ and low by a factor of about 2.0 for $\Xi^- + \overline{\Xi}^-$ at both 900 GeV and 7 TeV (see Fig. 11). Figure 12 shows that while Tune Z1 produces roughly the correct overall yield of kaons, the $p_T$ dependence of the produced kaons is too "soft". The data show more kaons at high $p_T$ ($p_T > 1.0$ GeV/c) than predicted by Tune Z1. As can also be seen in Figure 12, the data also show more protons and antiprotons at large $p_T$ than predicted by Tune Z1. However, the overall yield of protons and antiprotons is too large for PYTHIA Tune Z1.

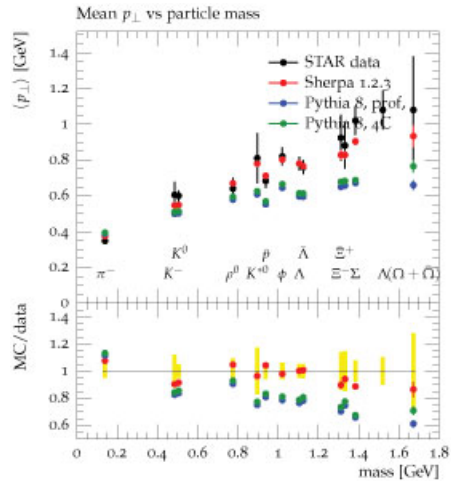

Figure 13: STAR data on the average $p_T$ versus mass compared with SHERPA and PYTHIA 8 from Hendrik Hoeth, http://users.hepforge.org/~hoeth/STAR_2006_S6860818/.

In summary, for PYTHIA Tune Z1 the overall yield of charged particles is slightly too low at both 900 GeV and 7 TeV. The overall yield of kaons is about right, hence the kaon to pion ratio is a bit high. The overall yield of protons and antiprotons is too high, while the overall yield of strange baryons is way too low. Also, PYTHIA Tune Z1 is off on the $p_T$ dependence of heavy particles, kaons, protons, lambdas, and cascades. The heavy particles have a harder $p_T$ spectrum that predicted by PYTHIA. We have known for a long time that the average $p_T$ increases with the mass of the produced particle. This can be seen clearly in the STAR data from RHIC in Figure 13 (from Hendrik Hoeth). We see that SHERPA [17] does a better job in describing the increase in average $p_T$ versus mass than does PYTHIA 8 [18].

## 4. SUMMARY & CONCLUSIONS

The PYTHIA 6.2 Tune DW which was created from CDF UE studies at the Tevatron did a fairly good job in predicting the LHC UE data 900 GeV and 7 TeV. The behavior of the UE at the LHC is roughly what we expected. The LHC PYTHIA 6.4 Tune Z1 does a very nice job describing both the CMS and ATLAS UE data at 900 GeV and 7 TeV. The UE is part of a hard scattering process. MB collisions quite often contain no hard scattering and are therefore more difficult to model. Since PYTHIA regulates both the primary hard scattering and the MPI with the same cut-off, $p_{T0}$, with PYTHIA one can model the overall "non-diffractive" (ND) cross section by simply letting the transverse momentum of the primary hard scattering go to zero. In this approach the UE in a hard-scattering process is related to MB collisions, but they are not the same. Of course, to model MB collisions one must also add a model of single (SD) and double diffraction (DD). Tune Z1 does a fairly good job of simultaneously describing charged particle production in both MB. I think it is amazing that it does as well on MB as it does.

Baryon production in MB collisions is a whole different story. For PYTHIA Tune Z1 the predicted overall yield of protons and antiprotons is too high, while the overall yield of strange baryons is way too low. Also, PYTHIA is off on the $p_T$ dependence of heavy particles, kaons, protons, lambdas, and cascades. The heavy particles have a harder $p_T$ spectrum that predicted by PYTHIA. It may be that other fragmentation schemes like SHERPA and HERWIG++ [19] might do better here than PYTHIA. SHERPA seems to do a better job in describing the increase in average $p_T$ versus mass than does PYTHIA 8. Remember, however, that the fragmentation models are constrained by the LEP $e^+e^-$ data and we do not want to have separate fragmentation tunes for $e^+e^-$ and hadron-hadron collisions! On the other hand, gluon fragmentation is not very well constrained by LEP and the hadron-hadron collider environment is much different than $e^+e^-$. Hadron-hadron collisions have BBR and MPI which are not present in $e^+e^-$ annihilations. For example, Figure 14 shows the pseudo-rapidity distribution for protons and lambdas at 7 TeV from PYTHIA Tune Z1. One can see the effect of the BBR at large $|\eta|$ values. However, the BBR have little effect in the central region where the data exist.

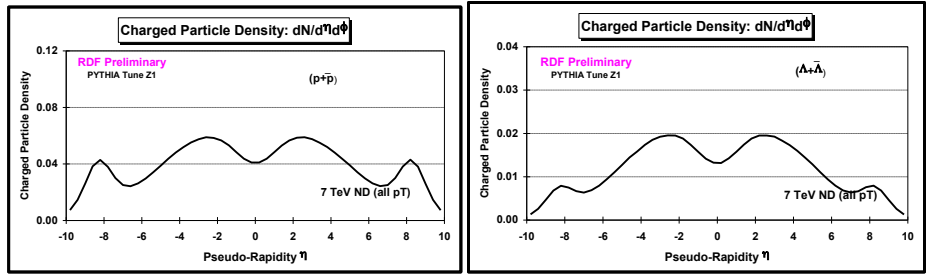

Figure 14: Shows the 7 TeV ND pseudo-rapidity distribution, dN/d$\eta$, for $p + \bar{p}$ and $\Lambda + \bar{\Lambda}$ from PYTHIA 6.4 Tune Z1.

In a very short time the experiments at the LHC have collected a large amount of data that can be used to study MB collisions and the UE in a hard scattering process in great detail. This data can be compared with the Tevatron MB and UE data to further constrain and improve the QCD Monte-Carlo models we use to simulate hadron-hadron collision. Both PYTHIA 6.2 Tune DW and PYTHIA 6.4 Tune Z1 describe the Drell-Yan UE data at the Tevatron (1.96 TeV) and the LHC (7 TeV). However, at present none of the tunes simultaneously describe perfectly the "jet" UE data at 900 GeV, 1.96 TeV, and 7 TeV. I do not know why the QCD models describe the Drell-Yan UE data better than they do the "jet" UE data. Too bad we do not have Drell-Yan UE data at 900 GeV. None of the Monte-Carlo models describe all the features of the MB data. I believe the tunes will continue to improve. We are just getting started! The future will include more comparisons with PYTHIA 8, SHERPA, and HERWIG++.